# Comments on
## "Experimental study of radiative shocks at PALS facility",
by Chantal Stehlé, Matthias González, Michaela Kozlova, Bedrich Rus, Tomas Mocek, Ouali Acef, Jean Philippe Colombier, Thierry Lanz, Norbert Champion, Krzysztof Jakubczak, Jiri Polan, Patrice Barroso, Daniel Bauduin, Edouard Audit, Jan Dostal and Michal Stupka
published in   Laser and Particle Beams, **28**, 253 (2010)

Michel Busquet, ARTEP inc, Ellicott City, MD, USA

**Abstract:**
Discussion on paper "Experimental study of radiative shocks at PALS facility" by C.Stehlé et al is presented, findings are questioned.

## I- introduction

In "Experimental study of radiative shocks at PALS facility" (Stehlé et al., 2010), (called hereafter paper I), C.Stehlé et al report results obtained on the PALS laser facility in 2007 to study the radiative precursor of a supercritical shock launched in a miniature shock tube, where the piston was accelerated by laser ablation from a 150 J, 0.35ns laser beam at $\lambda=0.438\mu m$.

Three different diagnostics are presented in paper I, time integrated XUV emission spectroscopy, time resolved interferometric measure of electron density along the direction of shock propagation, and interferometric measure perpendicular to this direction. This probe was made possible through specially designed miniature tube shocks (Busquet, 2010a). Analysis of all three diagnostics, as presented in paper I, needs some corrections that I will discuss in the following sections. I shall also give additional details from this experiment to which I participated as the second P.I.

I will not recall here the details of the experimental setup, as readers will find them in paper I.

## II- About XUV Emission Spectroscopy

As the sensitivity of the XUV emissivity of the Xenon column of gas was not large enough for the sensitivity of spectrograph in use, its entrance slit had to be removed. The resulting very poor spectral resolution was then given by the transverse dimension (700μm) of the source and the whole section of the tube was probed, *not the central region* as written in paper I. A reference spectrum (this one obtained with a thin slit) emitted by the alumina covering a solid aluminium foil directly heated by the laser is presented in Fig.1 and compared to 3 of the 4 spectra obtained from "blind targets".

This comparison clearly demonstrates that the recorded spectra are dominated by the Oxygen lines emitted by the alumina coating the aluminium walls, or by the air dopant added to the Xenon gas (bottom spectrum). The latter case is the brighter one, which confirms the Oxygen origin of the observed spectrum. Recorded levels range from 150 (for a 4th spectrum not shown here) to 7800 counts per pixel. These spectra cannot be compared to, *and cannot*

*be claimed to be in qualitative agreement* with numerical predictions computed for pure Xenon.

Note also that the wavelength calibration presented in paper I is erroneous, the wavelength range in Fig.7 of paper I should read 14 to 20.5 nm instead of 12 to 23 nm.

The targets with windows were designed (Busquet, 2010a) so that some small step of solid Al hides the wall emission. There is no need for the "development of some plasma from the end of the tube" to explain absence of record on the XUV spectrometer with these targets, as written in Sec.3.1 of paper I.

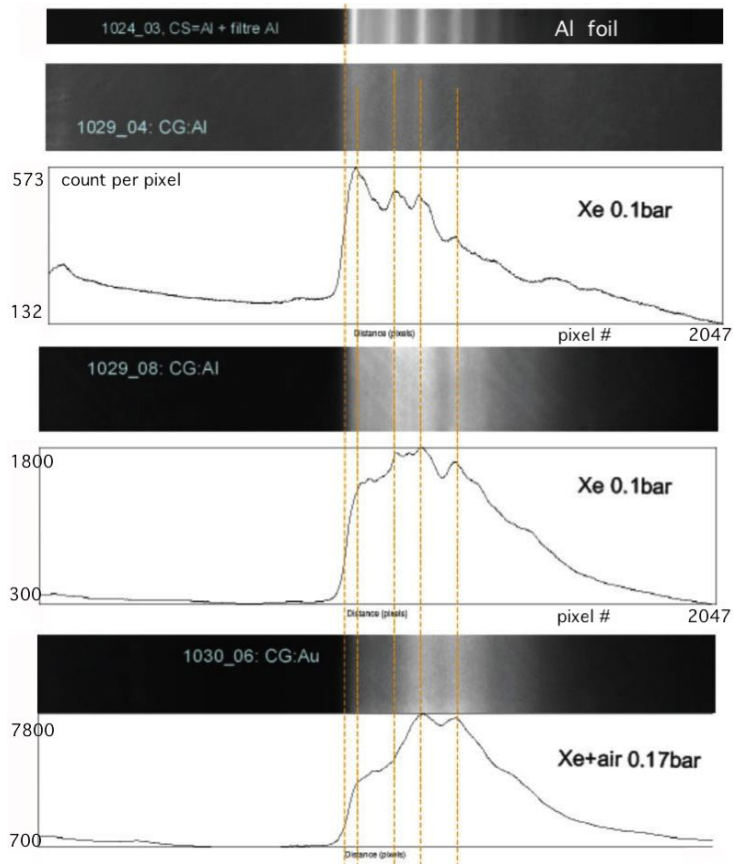

*Fig.1 : recorded spectrum from alumina, using a solid aluminium foil (upper image) compared to 3 spectra from "blind targets" (resp. CG35, CG32 & CG50). The spectrograph entrance slit has been removed for those 3. The spectral range is approximately 9-26 nm. Poor reproducibility (as between spectra 2 and 3) has been found.*

### III- About interferometric probe parallel to the tube axis

This is the central point of paper I. The authors claim that stationary limit has been reached, i.e. that the precursor velocity becomes equal to the shock velocity, or equivalently that the distance D between the shock (the bulk density jump) and the radiative precursor front (the electronic density rise, following the temperature and ionization rise) is no more varying with time. However the *shock location cannot be inferred* from the recorded interferogram and certainly *cannot be identified to the grey line* of Fig.3 of paper I. In fact, the presence of fringes on the lower left of the figure proves that the piston has not yet reached the position of these fringes, and thus has a velocity below 40 km/s. The shock velocity is probably not much larger, and then much smaller that the observed precursor front velocity (65 km/s). Moreover, crossing of the precursor front with reverse waves coming from the back of the tube (Fig.2) *prevents any analysis* of the precursor front after 30 ns.

Therefore the experiment *does not* "suggest that a quasi-stationary limit has been approached" as claimed in the conclusion of paper I.

On the other hand, as one can see in Fig. 5 of paper I displaying density profiles computed with the hydrodynamic code HERACLES, the theoretical value of D is still increasing, from 0.55 mm at 20 ns to 0.65 mm at 40 ns.

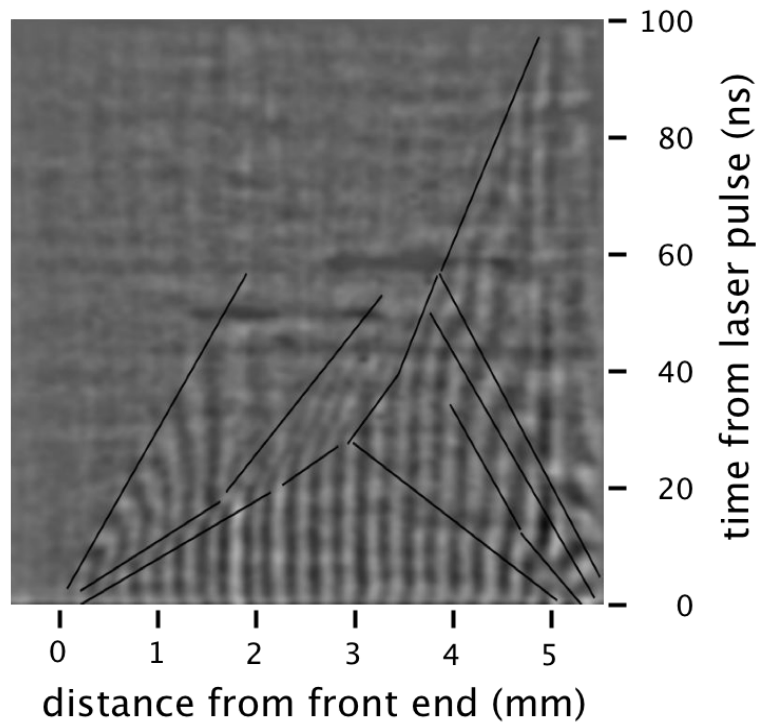

*Fig.2 : Limits of zones that can be identified in the interferogram of shot #1029_06 . The horizontal axis is the distance (0-6 mm) from the initial position of the piston, the vertical axis is the elapsed time (0-100ns) since the laser pulse arrival. The image has been processed by background subtraction followed by a FFT smoothing. Horizontal grey traces at 50 and 60ns are burnt pixels of the CCD camera.*

**IV- About interferometric probe perpendicular to the tube axis**

In the shot#1101-02 analyzed in paper I, some mask in the optical path reduced by 0.2 mm the field of view of the recorded interferogram (on the left of Fig. 6 of paper I), therefore the focal spot was off-center by less than 50 μm, compared to the 1.2 mm width of the channel (Busquet , 2010b). In this section the word "shock" used by the authors is to be understood as "precursor front".

## V- Miscellaneous

Several glitches in the description of the experimental setup may have obscured the readers. Obviously the wavelength of the third harmonic is to read $\lambda_3 = \lambda_1/3$, not $\lambda_3 = 3/\lambda_1$. The interferometric band-pass filter was set at the probe wavelength $\lambda_P$=527nm, not at the main pulse wavelength $\lambda_3$ (=438nm). The hiccup of the fringes at the laser arrival time is not related to any preheating and anomalous refractive index, but is merely the trace of an electromagnetic perturbation to the streak camera electronics following the intense and rapidly rising electric currents and magnetic fields created by the gas breakdown.

## VI- Conclusion

In conclusion, major and minor points claimed by paper 1 were not experimentally founded. No approach to stationary limit has been observed. Observation of Xenon emission cannot be ascertained, and cannot be compared to theoretical spectrum. The paper also suffers from many inaccuracies. Finally the electronic density profiles from three different shots have been carefully derived elsewhere and can be found in (Busquet, 2010b).

## VII- References